\documentclass[11pt]{article}

\usepackage[utf8]{inputenc}
\usepackage[T1]{fontenc}
\usepackage{authblk}
\usepackage[margin=1in]{geometry}
\usepackage{cite}
\usepackage{xurl}
\usepackage{multicol}

\title{\textbf{Beyond Explainability: The Case for AI Validation}}

\author{
\begin{multicols}{2}
\begin{tabular}{c c}
\begin{tabular}{c}
\textbf{Dalit Ken-Dror Feldman} \\
University of Haifa \\
\texttt{dalitkd@law.haifa.ac.il}
\end{tabular}
\begin{tabular}{c}
\textbf{Daniel Benoliel} \\
University of Haifa \\
\texttt{dbenolie@law.haifa.ac.il}
\end{tabular}
\end{tabular}
\end{multicols}
}

\date{May 2025}

\begin{document}
\maketitle
\begin{multicols}{2}
\begin{abstract}
Artificial Knowledge (AK) systems are transforming decision-making across critical domains such as healthcare, finance, and criminal justice. However, their growing opacity presents governance challenges that current regulatory approaches, focused predominantly on explainability, fail to address adequately. This article argues for a shift toward validation as a central regulatory pillar. Validation, ensuring the reliability, consistency, and robustness of AI outputs, offers a more practical, scalable, and risk-sensitive alternative to explainability, particularly in high-stakes contexts where interpretability may be technically or economically unfeasible. We introduce a typology based on two axes, validity and explainability, classifying AK systems into four categories and exposing the trade-offs between interpretability and output reliability. Drawing on comparative analysis of regulatory approaches in the EU, US, UK, and China, we show how validation can enhance societal trust, fairness, and safety even where explainability is limited. We propose a forward-looking policy framework centered on pre- and post-deployment validation, third-party auditing, harmonized standards, and liability incentives. This framework balances innovation with accountability and provides a governance roadmap for responsibly integrating opaque, high-performing AK systems into society.
\end{abstract}


\section{Introduction}
The rise of Artificial Knowledge (AK) challenges traditional models of trust, accountability, and governance due to its opaque nature. While AK systems can process vast datasets and generate powerful insights their complexity often prevents full understanding, introducing epistemic uncertainties~\cite{rajpurkar2022ai,wu2019deep,mckinney2020evaluation}.  This becomes a critical concern in fields like healthcare~\cite{cross2024bias}, finance~\cite{li2021algorithmic,lawton2022ai}, and criminal justice~\cite{ccj2024ai,rigano2019ai,zavrsnik2020ai,sushina2019ai}, where unverified outputs impact fundamental rights. This paper advocates shifting the focus from explainability alone to a combination of explainability and validation of AK systems, offering a more practical and scalable approach for integrating AK into high-stakes domains while preserving trustworthiness and safety.

\section{The Problem with Explainability}
The primary focus of this article is one of the most critical challenges facing AK: its inherent opacity~\cite{oecd2019,euaiact2021_13_25,gdpr2016}.  Artificial systems mitigate this issue by embedding explainability into their outputs, enabling trust. Explainability is grounded in the notion that it promotes trustworthy AI applications~\cite{baker2023explainable}.  However, balancing explainability with technical accuracy presents challenges, as simplifying complex models may compromise precision, raising concerns about fairness and accountability~\cite{yampolskiy2019}. Explainable Artificial Intelligence (XAI) aims to enhance the transparency and trustworthiness of complex "black-box" models by offering human-interpretable explanations for their outputs. This is particularly critical in high-stakes domains where AI-driven decisions require both accountability and user trust. However, recent literature reveals fundamental problems: the notion of interpretability is often ambiguous~\cite{lipton2018mythos,rozen2023case},  and many explanation methods are unreliable or misleading~\cite{baker2025monitoring}.  Poor-quality explanations may create only an illusion of understanding, leading to incorrect human decisions, overconfidence in flawed models, or even enabling manipulation~\cite{martens2025beware}.  
Explanations can obscure model behavior, as seen in the "white box paradox", where they increase trust despite incorrect recommendations. Efforts to clarify black-box models may falsely create transparency, undermining usability and introducing errors~\cite{cabitza2023painting}.   Some take it to the next level and call it the 'XAI halo effect'. This occurs when misleading explanations lead users to trust advice without verifying its accuracy~\cite{cabitza2024explanations}.  Thus, misleading explanations can diminish the benefits of transparency. Moreover, relying on explanations rather than designing inherently interpretable models may perpetuate harmful practices~\cite{rudin2019stop,lipton2018mythos}.

\section{The Advantages of Validation}
In the absence of full explainability tools~\cite{rozen2023case,rozen2024lost,ghassemi2021false}, validated data can help close the regulatory gap, with the system's overall utility could compensate for the limitations inherent in its lack of explainability~\cite{nicholls2024pentesting,nidhra2012testing,khan2012comparative}.  
Validation forms the objective foundation for fairness initiatives by ensuring consistent and reliable system outputs. Both the EU AI Act (article 10)\cite{euaiact2021} and the U.S. Algorithmic Accountability Act of 2022\cite{aaa2022} emphasize fairness, which in contexts like hiring platforms under the GDPR, requires validated processes to prevent discriminatory automated decisions~\cite{fabris2025fairness,rigotti2024fairness,gdpr2016}. Validation focuses on outcomes rather than processes, making it a more feasible and scalable solution for ensuring trust in complex AI systems~\cite{myllyaho2021validation,fda2024devices}. Validation outweighs efficiency. Prioritizing efficiency over validity can result in significant risks, if a system is not adequately validated~\cite{microsoft2023azure,dau2025asoe}. 
Validation ensures efficiency does not compromise reliability. For instance, the EU's AI Act (article 16)~\cite{euaiact2021} and the U.S. Federal Reserve’s Guidance on Model Risk Management~\cite{federalreserve2011} both mandate testing and validation of high-risk AI systems to ensure robustness, prioritizing long-term stability over short-term efficiency gains. By requiring validation, policymakers establish clear benchmarks for assessing AK reliability. For example, China’s AI regulations require validation reports for high-risk systems, giving regulators evidence to assess compliance (articles 7-8, 2022)(article 17, 2023)~\cite{cac2022,cac2023}.  
Moreover, validation inherently ensures safety, especially in high-stakes areas like healthcare and public safety. The FDA's Good Machine Learning Practices (GMLP)~\cite{fda2021gmlp}, and the EU's AI Act (article 10) mandate rigorous testing to ensure systems operate reliably under diverse conditions. For example, autonomous vehicles undergo simulations and real-world tests, where safety is integral to the validation process, not a separate concern~\cite{eastman2023vehicles,ebert2019validation,anderson2016autonomous}. 
Lastly, validation is crucial for ensuring AI systems perform as intended, offering an objective standard for reliability while supporting values like fairness, accountability, and safety. Frameworks such as the EU's AI Act, the FDA's GMLP, and China's AI regulation emphasize validation as central to AI governance, addressing risks, fostering trust, and ensuring compliance with regulatory and societal priorities.

\section{Discussion: The Validity-Explainability Matrix}

Balancing validity and explainability in AI systems is challenging. Highly explainable systems that lack validity may gain trust while producing unreliable outputs, while valid systems with low explainability can erode trust and accountability. This section explores the trade-offs between validation and explainability. It critiques the current focus on explainability, emphasizing that validity and explainability should be complementary. 
Validation ensures output reliability and consistency, while explainability makes outputs interpretable. Valid-explainable AK achieves reliability consistency and interpretability, but is often limited to simpler models or low-stakes contexts where explainability costs are manageable, such as linear regression models in inventory management~\cite{hoppenheit2015inventory}. 
In contrast, valid-opaque AK is reliable and consistent but lack explainability~\cite{messenger2004minoxidil}. This category often includes deep learning systems, where outputs are generated through non-linear and probabilistic processes~\cite{sahin2024blackbox}.  While these AI systems may achieve high levels of performance, the costs of making them explainabl, through post hoc interpretability tools~\cite{turbe2023posthoc}, documentation, or causal mapping~\cite{carloni2023causality}, can be excessive. Achieving explainability of valid-opaque AK may reduce model’s accuracy, delay deployment, or impose significant technical and financial burdens. This trade-off is particularly pronounced in domains where performance optimization is paramount, such as financial risk modeling or AI-driven supply chains~\cite{kosasih2024supplychain,finance2024risk}. 
Non-valid opaque AK exacerbates these issues. It lacks reliability, consistency, and explainability, imposing the highest risks and costs on society. The opacity of such AI systems makes it difficult to detect errors, while their invalid outputs can propagate cascading failures. For instance, a flawed AI system used in credit scoring may penalize certain demographic groups based on incomplete or biased data. Without explainability, stakeholders cannot identify or challenge the source of these biases, leaving societal harm unaddressed~\cite{ghassemi2021false}. 
Given these dynamics, validation is a more practical and cost-effective approach when explainability is not feasible. It ensures consistent, accurate outputs, even if the internal processes are unclear. In high-stakes areas like healthcare or finance, reliable results are crucial for minimizing risks. For instance, a validated medical diagnostic tool provides trustworthy results despite lacking explainability~\cite{fassi2024validation,ghassemi2021false}.  
Balancing validity and explainability in AI systems presents significant challenges. Highly explainable systems that lack validity may gain trust but produce unreliable outputs, while valid but opaque systems can undermine trust and accountability. The EU, U.S., and China have different approaches to balancing these factors, but validation remains crucial, especially in high-risk areas like healthcare and public safety (EU's AI Act, article 13(3)(b)(ii))~\cite{euaiact2021}. 
It is worth noting that the EU’s Framework Convention on AI of 2024~\cite{coe2024framework}, does not explicitly reference the terms “explainability” or “validity”, perhaps reflecting a growing skepticism about the feasibility of fully explaining AI systems. Nevertheless, the document emphasizes the principles of oversight and transparency (article 8)~\cite{coe2024framework} accountability, responsibility (article 9)~\cite{coe2024framework} and reliability (article 12)~\cite{coe2024framework}. It highlights the importance of maintaining human control over AI processes, even in the absence of a detailed understanding of their mechanics. Comparably, in the UK White Paper, the concept of explainability is mentioned, but the requirement for validity is not addressed~\cite{uk2023approach}. This arguably leaves room for greater emphasis on validation as the more practical alternative in specific contexts.
In contrast, U.S. policy places greater emphasis on validity. In 2023, President Biden issued the Executive Order on the Safe, Secure, and Trustworthy Development and Use of AI, mandating robust evaluation measures to ensure AI systems are reliable, secure, and function as intended (section 2(a))~\cite{execorder14960_2023}. This comprehensive framework governs the development, testing, and deployment of AI technologies (section 4.1(a))~\cite{execorder14960_2023}. The Executive Order emphasizes the importance of validity in AI systems, particularly through its focus on safety, security, and trustworthiness (sections 4.2 and 4.3)~\cite{execorder14960_2023}. Although this Executive Order was canceled by President Trump in 2025~\cite{trump2025eo14179}, the notion of validity remained. The Office of Management and Budget (2025)~\cite{omb2025m2521} highlights the importance of continuous validation of AI systems. Similarly, the 2020 executive order by President Trump~\cite{trump2020trustworthy}, emphasizes the need for AI systems to be "accurate, reliable, and effective" in their deployment, further reinforcing the need for validation in AI applications.
Both validity and explainability are included in the proposed California legislation~\cite{california2023ab331}, as well as in the National Institute of Standards and Technology (NIST) Principles~\cite{nist2024ai6001}, reflecting a broader emphasis on these values for ensuring the reliability, responsibility and trustworthiness of AI systems. The emphasis on validation, though, must not become a justification for perpetuating opacity~\cite{myllyaho2021validation}. Developers should be required to invest in explainability where its costs are reasonable and its benefits, such as fairness, accountability, and human oversight, are significant~\cite{fabris2025fairness,rigotti2024fairness}.  
Policymakers must balance the incentives for validation and explainability through targeted regulation and economic mechanisms. Pre-deployment validation mandates, such as those in the EU AI Act, provide a baseline standard of reliability. Liability frameworks, such as strict liability for system failures, ensure that developers internalize the costs of harm, incentivizing robust testing even when explainability remains elusive. 
AK can be classified along explainability and validity axes, highlighting the trade-offs and challenges in its governance. Table 1 below provides a structured framework for this analysis.

\begin{table*}[t]
\centering
\begin{tabular}{|c|c|c|}
\hline
\textbf{} & \textbf{Explainable} & \textbf{Opaque} \\
\hline
\textbf{Valid} & Valid-Explainable & Valid-Opaque \\
\hline
\textbf{Non-Valid} & Pre-Valid Explainable & Non-Valid Opaque \\
\hline
\end{tabular}
\caption{Typology of Artificial Knowledge Creation}
\label{tab:ak_typology}
\end{table*}

This typology underscores the balance required between explainability and validity, offering a structured approach for governance and regulatory strategies to maximize AK’s benefits while minimizing risks.

\subsection{Valid-Explainable: Reliable, Consistent, Interpretable, and Transparent}
Balancing \textit{validation} and \textit{explainability} in AI presents challenges. While explainable AI can build trust, its complexity can increase costs, and it may not always improve reliability. Validation, however, ensures AI systems perform as intended, making it crucial in high-stakes domains where accuracy is vital. Regulatory frameworks like the EU AI Act, U.S. executive orders, and China’s AI regulations all emphasize validation but differ in their treatment of explainability. Validation, often more cost-effective than explainability, reduces risks, inefficiencies, and societal harm, which can occur from unreliable systems.
In sectors like healthcare, validated systems are more likely to be trusted and adopted. However, this focus on validation shouldn’t justify ignoring explainability where it's feasible, as explainability ensures fairness, accountability, and oversight. Policymakers must balance validation and explainability through targeted regulations, liability frameworks, and market incentives like certification programs, which encourage innovation while maintaining reliability. Ultimately, in some cases, validation may be the more practical approach, fostering trust and safeguarding societal interests when explainability is economically unfeasible. Recent research explores hybrid models that balance both aspects, aiming for high accuracy while maintaining interpretability. For instance, partially interpretable models provide explanations for decisions while ensuring robust performance~\cite{frost2024interpretable,gramegna2021shaplime,slack2020fooling}.   

\subsection{Pre-Valid Explainable: Interpretable but Provisional}
The \textit{Pre-Valid Explainable} quadrant represents AK that prioritizes explainability but has not yet achieved full validation. This knowledge is transparent, interpretable and accessible, allowing stakeholders to understand AI outputs and reasoning, but their findings remain provisional. That is due to limited testing, incomplete training, or insufficient contextual robustness~\cite{chander2024tai}. While unsuitable for high-stakes applications, these systems may be indispensable in exploration, developmental, and speculative contexts.
This approach is particularly fitting for prevalid explainable systems. While interpretable and transparent, these systems are provisional and prone to errors. Traditional regulatory models may prematurely dismiss or overburden them. Experimentalist governance offers a flexible pathway for refinement, with regulatory sandboxes providing a controlled environment to test prevalid systems with reduced compliance requirements~\cite{ranchordas2021experimental}. These sandboxes allow regulators and developers to evaluate system performance, identify risks, and implement improvements without the pressure of full-scale deployment~\cite{ranchordas2021experimental}. 

\subsection{Valid-Opaque: Reliable and Consistent but Inscrutable}
The \textit{Valid-Opaque} quadrant characterizes AK, demonstrating high reliability and consistency but a lack of explainability. These systems consistently produce outputs that meet rigorous standards of accuracy and robustness, yet their internal decision-making processes remain opaque\cite{poursabzi2021interpretable,kaur2020interpreting}. This combination of validity and opacity presents significant challenges in trust, accountability, and governance\cite{kaur2020interpreting,poursabzi2021interpretable}. That is, especially in high-stakes applications where decisions have profound societal implications\cite{kaur2020interpreting, poursabzi2021interpretable,ghassemi2021false}. 
Opacity in valid-opaque systems stem from complex architectures like neural networks and ensemble methods, which are difficult to interpret despite their effectiveness\cite{elaziz2021metaheuristic,aziz2018ann}. This lack of explainability creates accountability challenges, making it hard to identify the sources of errors or biases\cite{mikalef2022darkside,fuchs2018bias}. Pasquale calls this the "accountability gap", where stakeholders can't discern whether issues arise from data, algorithms, or external factors\cite{pasquale2015blackbox}.  
However, balancing validity and opacity requires careful governance. Policymakers must navigate this tension through accountability frameworks, transparency incentives,\cite{denzel2024certification} and liability regulations\cite{europeancommission2020safety, europeancommission2019liability,bertolini2021liability} to ensure responsible and ethical deployment of these systems\cite{europeancommission2020safety}, prioritizing trust, fairness, and accountability\cite{myers2020unreliable}.

\subsection{Non-Valid Opaque: Unreliable and High-Risk}
The \textit{Non-Valid Opaque} quadrant represents the most concerning and problematic category of AK: artificial intelligence systems that produce unreliable outputs while remaining inscrutable.  These outputs fail to meet empirical standards of accuracy, consistency, and robustness, and their opacity prevents stakeholders from diagnosing or addressing underlying flaws\cite{pfau2024trustworthy,banerjee2020failures}. This combination of unreliability and opacity exacerbates the externalities of AK, undermining trust, equity, and the efficiency of broader knowledge systems. 
Non-valid opaque systems often arise from governance failures, such as poor-quality training data, inadequate validation, and misaligned incentives\cite{dakka2021dataquality}. These systems, developed under market pressures or weak oversight\cite{kancharla2019speed}, can perpetuate biases and errors. Unlike valid-opaque systems, which can be mitigated with interpretability tools, non-valid opaque systems require proactive controls, such as pre-deployment audits and stress testing, to prevent harm\cite{mokander2024audit,sivathapandi2022synthetic}. In some cases, outright prohibition may be necessary to protect public safety and equity. These systems also erode public trust in AI, creating long-term societal costs, particularly for marginalized groups\cite{ccj2024criminal,taylor2023algorithm, carlson2017transparency}. 
The economic impact is also significant, as non-valid opaque systems lead to litigation\cite{jacobsen2023smart}, reputational damage\cite{soyer2022liability,holweg2022reputation}, and hinder innovation\cite{stiglitz2000econinfo}. They prevent iterative improvements, reinforcing a cycle of unreliability\cite{baek2024researchagent}. Thus, addressing non-valid opaque systems requires robust regulation, accountability frameworks, and, in some cases, prohibition, to ensure that AI development aligns with societal values of trust, equity, and innovation.

\section{Conclusions and Future Research}
In conclusion, integrating AK systems into high-stakes domains introduces significant trust, accountability, and governance challenges. While explainability is essential, the focus should also shift toward validating the reliability and consistency of these systems' outputs. Relying on explanations rather than designing inherently interpretable models may perpetuate harmful practices; however, validation, which ensures consistency and accuracy over time, can address these issues by providing a more reliable foundation for trust and decision-making. This approach offers a practical pathway for adopting AK in healthcare, finance, and criminal justice without undermining societal values. 
To effectively govern AK systems, a comprehensive, risk-based regulatory framework is required to integrate legal, economic, and institutional mechanisms. Such a framework should mandate both pre-deployment and post-deployment validation. Such a framework would establish independent third-party bodies for the impartial evaluation of high-risk systems and support the development of publicly funded validation infrastructure. The latter include standardized datasets, data repositories, and computational resources, particularly to ensure access for small and medium enterprises\cite{denny2019allofus}.   
Importantly, validation should encompass functional, empirical, and normative dimensions, thereby ensuring technical robustness, fairness, and legal compliance. These mechanisms should be complemented by fallback procedures, such as human oversight in sensitive applications, and transparency requirements, including documentation of data sources, methodologies, and system limitations. Harmonization of validation standards through international bodies such as ISO and the OECD\cite{iso2022ai23053, oecd2024ai}, alongside the standardization of fairness metrics and tools, will further reduce compliance burdens and enhance global trust. Additional measures, liability regimes, certification schemes, public-private partnerships for fairness audits, and mandatory fairness and ethical impact assessments, can further align innovation with accountability. Together, these measures create a comprehensive policy framework that aligns innovation with accountability, ensuring that AK systems are reliable, fair, and trustworthy across diverse contexts.
Future research should explore methods for adequate validation of AK systems, particularly in complex, high-risk contexts. Additionally, developing hybrid models that balance explainability and performance and address AK's governance implications will be critical for ensuring these systems' responsible and ethical deployment. Further work is required to develop and refine the regulatory and policy recommendations outlined above, establishing a robust governance framework for AK systems that balances innovation with accountability, safeguards public interests, and fosters societal trust. Similarly, it is essential to explore the limitations of validation in dynamic systems and understand to what extent validity can truly be assessed.

\section*{Author’s Note}
This article is based on the theoretical framework first introduced in Daniel Benoliel, Dalit Ken-Dror Feldman. The social cost of artificial knowledge. University of Illinois Journal of Law, Technology \& Policy, 2025. Forthcoming.

\bibliographystyle{unsrt}
\bibliography{references}
\end{multicols}
\end{document}